\documentclass[10pt,twocolumn,twoside,submit]{JCNtran}
\usepackage{psfig,subfigure,array,amssymb,latexsym,amssymb,epsf,amsmath,cite} 

\usepackage[dvips]{epsfig}
\usepackage[latin1]{inputenc}
\usepackage[T1]{fontenc}

\def\BibTeX{{\rm B\kern-.05em{\sc i\kern-.025em b}\kern-.08em
    T\kern-.1667em\lower.7ex\hbox{E}\kern-.125emX}}

\begin{document}
\bibliographystyle{jcn}

\title {Optimization of Unequal Error Protection Rateless Codes for Multimedia Multicasting}

\author{Yu~Cao, Steven~D.~Blostein and~Wai-Yip~Chan 
\thanks{Manuscript received May 07, 2014; approved for publication by Prof. Emanuele Viterbo, Editor, December 13, 2014.}
\thanks{This research is supported by the Natural Sciences Engineering and Research Council of Canada Grant STPSC 356826-07, and by Discovery Grant 41731.}
\thanks{Y. Cao is with Huawei Technologies Canada, Kanata, Ontario, Canada K2K 3J1, {yucaoece@gmail.com}, S.D. Blostein and W.-Y. Chan are with Department of Electrical and Computer Engineering, Queen's University, Kingston, Ontario, Canada, K7L 3N6, \{steven.blostein, chan\}@queensu.ca.}
\thanks{The results of the paper are presented in part in \cite{Cao-isit-10} and \cite{Cao-bmsb-10}.}}
\markboth{JOURNAL OF
COMMUNICATIONS AND NETWORKS, VOL. XX, NO. YY, MONTH
2015}{Cao \lowercase{\textit{et al}}.: Optimization of Unequal Error Protection Rateless...} 
\maketitle

\begin{abstract}
Rateless codes have been shown to be able to provide greater flexibility and efficiency than fixed-rate codes for multicast applications. In the following, we optimize rateless codes for unequal error protection (UEP) for multimedia multicasting to a set of heterogeneous users. The proposed designs have the objectives of  providing either guaranteed or best-effort quality of service (QoS). A randomly interleaved rateless encoder is proposed whereby users only need to decode symbols up to their own QoS level. The proposed coder is optimized based on measured transmission properties of standardized raptor codes over wireless channels. It is shown that a guaranteed QoS problem formulation can be transformed into a convex optimization problem, yielding a globally optimal solution.  Numerical results demonstrate that the proposed optimized random interleaved UEP rateless coder's performance compares favorably with that of other recently proposed UEP rateless codes.
\end{abstract}

\begin{keywords}
multimedia, error control coding, unequal error protection, raptor codes, video transmission
\end{keywords}

\section{Introduction}
\label{sec:intro}

Multimedia transmission is a main driver for explosive data traffic growth in wired and wireless networks. While decades of research have been conducted in designing reliable multimedia transmission over error-prone channels, multimedia multicast over lossy packet networks is still challenging due highly variable channel conditions among different users,  QoS constraints and multimedia devices, e.g., smart phones, tablets, laptops. Scalable video coding (SVC) \cite{schwarz-Tcircuits-2007} is useful as it layers the source to enable efficient progressive reconstruction at the receiver.

Protection against channel impairments can be achieved by using codes that provide forward error correction (FEC): Reed-Solomon  \cite{Mohr-JSAC-00}, low density parity check (LDPC)  \cite{Raja-EJASP-07}, Turbo  \cite{Jaspar-ProcIEEE-07} and fountain \cite{Xu-JSAC-07} \cite{Ahmad-TCSVC-10}, as well as  joint source-and-channel coding (JSCC) \cite{Raja-EJASP-07} \cite{Jaspar-ProcIEEE-07} \cite{Xu-JSAC-07}. Other approaches that exploit source scalability to provide UEP use hybrid automatic repeat request or cross-layer optimization \cite{Liu-JSAC-10} \cite{VdS-TMC-06} \cite{Maani-TSVT-10}. The above approaches, however, were mainly envisioned for point-to-point links and do not consider heterogeneous users'  QoS. As a result, adaptation of JSCC to multimedia multicast is often inefficient in that they cater to the lowest QoS user.

A critical aspect of robust multimedia multicast is channel-coding performance. Traditional fixed-rate FEC encounters the problem of channel heterogeneity as in the case of Reed-Solomon (RS) codes that are targeted for one specific loss rate \cite{Lin-04}. {\em Rateless} fountain codes \cite{mitzenmacher-itworkshop-04} are efficient and flexible for broadcasting or multicasting over erasure channels. The rateless property enables (1) a transmitter to generate, as needed, an unlimited number of encoded symbols, and (2) a receiver to successfully recover any subset of the encoded symbols of size slightly greater than the number of information symbols. Raptor codes \cite{shokrollahi-tit-06} due to their high performance and low complexity are fountain codes that have been incorporated into the third generation partnership program (3GPP) Multimedia Broadcast/Multicast Services (MBMS) standard  \cite{3GPP-09}.  In \cite{luby-Tbroadcast-07}, raptor codes have been extensively evaluated for MBMS download delivery.  A more recent version appears in  \cite{Luby-RaptorQ-11}, and background can be found in \cite{Luby-09}. Rateless codes have been applied to SVC-based multi-source streaming \cite{Schierl-WCM-06}, adaptive unicast streaming \cite{Ahmad-TCSVC-10}, and SVC streaming from multiple servers  \cite{Wagner-ICME-06}. A JSCC rateless coding framework for scalable video broadcast appears in \cite{Ji-TMM-12}. Applications to distributed video streaming for relay/cooperation based  receiver-driven layered multicasting is found in \cite{Xu-TCSVT-07}, while \cite{Golrezaei-Magzine-13} and \cite{Shanmugam-TIT-13} use fountain codes for distributed video caching via user cooperation.

While the  raptor code itself is not suited for progressive decoding, multimedia has a hierarchical source symbol priority structure necessitating unequal error protection (UEP), sometimes referred to as priority encoding transmission (PET) \cite{Albanese-Tit-96}. Numerous UEP approaches to multimedia transmission have been proposed  \cite{Mohr-JSAC-00} \cite{Chou-pvw-03} \cite{Hamzaoui-SPM-05} \cite{Stankovic-SP-05}. In \cite{Mohr-JSAC-00}, Mohr proposes a PET-based packetization scheme for transmitting compressed images over noisy channels. In \cite{Chou-pvw-03}, the Mohr scheme is optimized to minimize end-to-end distortion. Optimization of receiver-driven networks has also been investigated \cite{Chou-TMultimedia-01}. Rate-distortion-based optimization can be found in \cite{Chou-TMultimedia-06}. Rather than incorporate code performance into the optimization, these existing optimization approaches generally employ maximum distance separable (MDS) codes.  In this paper, code performance is taken into account in the UEP rateless code optimization.

Not surprisingly, UEP rateless code design methods have recently appeared. In \cite{Rahnavard-TIT-07}, message symbols are encoded by non-uniform selection of source symbols
and applied to MPEG-II video transmission in \cite{Talari-Milcom-09}. In \cite{Sejdinovic-Tcom-09}, expanding window fountain (EWF) codes organize source symbols into a sequence of nested windows. In \cite{Vukobratovic-Tmultimedia-09}, EWF codes are applied to scalable video multicasting. Windowing approaches for rateless codes that achieve equal error protection (EEP) have been proposed in \cite{Studlholme-ISIT-06} and \cite{Bogino-ISCAS-07}. The sliding-window (SW) rateless code design proposed in \cite{Bogino-ISCAS-07} is applied to wireless video broadcasting in \cite{Cataldi-TIP-10}. Ahmad et. al \cite{Ahmad-TMM-11} achieve UEP  in video multicast using the Luby Transform (LT) \cite{luby-focs-02} via block duplication. In \cite{Yang-Tcomputers-12}, a UEP rateless code based on hierarchical graph coding is proposed for media streaming. However, these previous UEP rateless code design approaches  \cite{Rahnavard-TIT-07} \cite{Sejdinovic-Tcom-09} may compromise performance as they alter the LT code \cite{luby-focs-02} degree distribution unless the degree distribution is jointly-optimized with UEP parameters.

Finally, previous approaches to UEP optimization for multimedia have focused almost exclusively on providing best-effort QoS, i.e., maximization of an average fidelity measure of video/image quality of end users for a given transmission rate
 \cite{Mohr-JSAC-00} \cite{Chou-pvw-03} \cite{Chou-TMultimedia-06} or with rateless codes \cite{Ji-TMM-12} \cite{Talari-Milcom-09} \cite{Vukobratovic-Tmultimedia-09} \cite{Cataldi-TIP-10}.  As rateless codes have no pre-determined transmission rate, QoS may be achieved by transmitting enough coded symbols to meet users' QoS demands.  In contrast, our focus is on guaranteed QoS optimization, i.e., minimizing resource usage under the constraints of heterogeneous QoS guarantees.  While \cite{Cao-isit-10} and \cite{Cao-bmsb-10} presented early versions of this approach, this paper provides more complete background, technical detail,  a  method to simplify constraints, as well as an example video multicasting application.

 Our main contributions are summarized as follows:
\begin{enumerate}
\item a UEP scheme is proposed that uses random interleaving of raptor coders that enables
direct application of already optimized standardized raptor codes used for 3GPP MBMS \cite{3GPP-09}.
When applied to multicasting to heterogenous users, low bandwidth clients need not receive encoded symbols targeted to high bandwidth clients, which can significantly reduce receiver complexity and time to decode.
\item the proposed design, optimized for multimedia multicast to heterogeneous users, contains QoS guarantees and factors in rateless code performance.
With  standardized raptor codes, this guaranteed QoS optimization problem is shown to be convex with a simplified solution using Karush-Kuhn-Tucker (KKT) optimality conditions.
\item through a combination of simulation and analysis, performance of the proposed random interleaved UEP rateless design is compared to other EEP  and UEP rateless coders.
\end{enumerate}

The paper is organized as follows: Section \ref{sec:sysmodel} describes the system setup and proposed UEP rateless code design; Section \ref{sec:problem formulation} presents the problem formulations for guaranteed and best-effort QoS; Section \ref{sec:solution} provides the solution for guaranteed QoS. Section \ref{sec:solution} transforms the original problem formulation for guaranteed QoS  into a convex optimization problem where optimal selection probabilities for interleaving are obtained in closed form for certain cases or else numerically. Comparisons with recent UEP rateless coding schemes are provided in Section \ref{sec:comparison}.

\section{System setup and proposed design}
\label{sec:sysmodel}

\subsection{System setup}
\label{subsec:syssetup}

A multimedia server that transmits multimedia content simultaneously to multiple users is considered, which may include streaming with strict delay requirements.
Multimedia content is divided into multiple coded blocks. The server first compresses each source block using a pre-defined source coder and then adds error protection to the source information using a rateless, e.g., raptor or LT code. Encoded symbols are then multicast over a wireless lossy packet network.

User subscribers are classified into $J$ classes according to reception capability. For Class $j$ users, reception capability $\delta_j$ is defined as the proportion of symbols that the receiver can successfully receive compared to the number of transmitted symbols, $1 \leq j \leq J$. Therefore, in each transmission session, the number of successfully received encoded symbols for each user in Class $j$ is $\delta_j M$, where $M$ is the number of symbols transmitted\footnote{For analytical simplicity, the number of received symbols for each user class is modeled as $\delta_j$ multiplied by the total transmitted as in \cite{Vukobratovic-Tmultimedia-09}.
}. Without loss of generality, we order the classes according to reception quality, i.e., $0< \delta_1 \leq \delta_2... \leq \delta_J \leq 1$. For example, a Class 1 user may represent a mobile cell phone with limited reception quality due to size and power restrictions, while a Class 2 user may represent an automobile equipped with larger antenna and higher bandwidth service. Reception capabilities are determined by channel quality and bandwidth between server and receiver, and no distinction is made between overhead incurred by symbol erasures and lost symbol transmission opportunities due to client bandwidth restrictions.  Users in different classes may also have different QoS requirements, and  \emph{outage QoS guarantees} are used to enable users to recover a given portion of source data with an achieved target probability. Without loss of generality, the term {\em peak signal-to-noise ratio} (PSNR), a common measure used for visual media quality, is used to denote QoS.

Let $K$ represent the number of information symbols in a raptor-coded source block. Assume the server transmits $M=(1+\varepsilon) K$ encoded symbols in order to meet all users' QoS demands,  where $\varepsilon$ is the total transmission overhead for all layers needed to combat losses of the heterogenous users in the multicast system.
For scalability, the coded source block is partitioned into $L$ layers in decreasing order of importance: Layer $1$ contains the most important symbols while Layer $L$ contains the least important symbols. For example, in video or image compression terminology,  Layer $1$ might represent the base layer (BL), and Layer $2$ the first enhancement layer (EL). The number of source symbols in Layer $l$ is denoted by $S_l, 1 \leq l \leq L$, and $K=\sum_{l=1}^L S_l$.

Successful decoding of layer $l$ requires layers $1, 2, \ldots , L-1$ to be decodable. Rather than jointly optimizing the source and channel coders, we focus on optimizing channel coding parameters for  a given source coder. Therefore, we assume that the values of $S_l , 1 \leq l \leq L$ are provided by a pre-determined scalable source coder.


\subsection{Proposed UEP rateless code}
\label{subsec:proposeddesign}

\begin{figure}[htb]
\begin{center}
\includegraphics[width=3.4in, keepaspectratio]{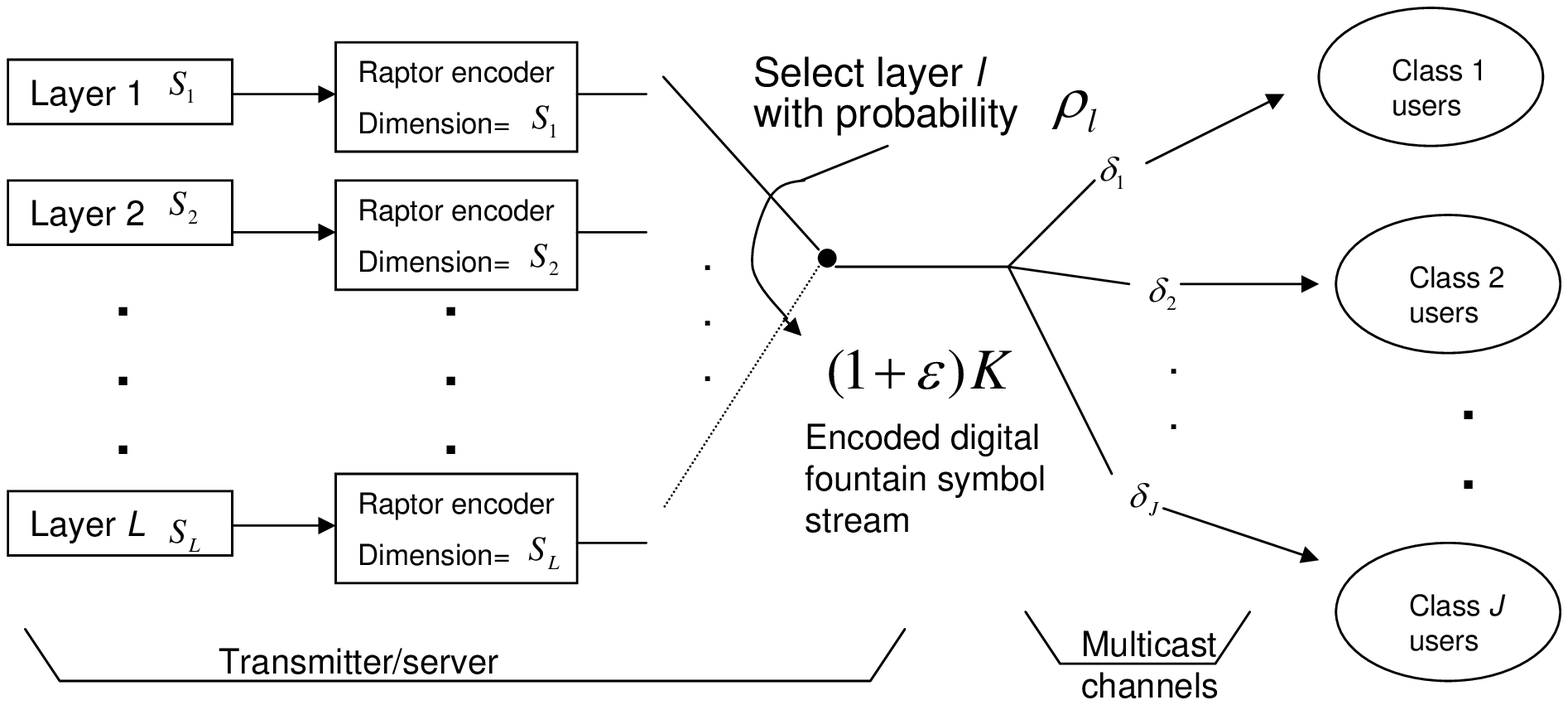} 
\caption{The proposed random interleaved UEP raptor coding method.} \label{Diagram}
\end{center}
\end{figure}

We propose a randomly interleaved UEP rateless encoder structure to provide FEC for multimedia multicast as shown in Fig. \ref{Diagram}. The encoder assumes that source symbols have been allocated to the $L$ layers prior to encoding. Encoding is performed by randomly selecting layer $l$ with probability $\rho_l$ for $l=1,2,...,L$ where $\sum_{l=1}^L \rho_l =1$. 
Encoded output symbols are generated by the raptor encoder for Layer $l$ with code dimension $S_l$, degree distribution $\Omega_l(x)$ and precode $C_l$. The overall encoded data stream consists of  interleaved raptor-encoded symbols from the $L$ encoders. From the above definitions, $\varepsilon$ can be lower bounded by
\begin{equation}\label{min OH}
\varepsilon_{\min} = \frac{1}{K} \sum_{i=1}^J \mbox{(Sel)}_i / \delta_i - 1,
\end{equation}
where $\mbox{(Sel)}_i \in \{ S_1, S_2, \cdots , S_L \}$ denotes the code dimension of the selected layer for user Class $i$.

The proposed rateless coded scheme uses the random interleaving to achieve UEP.
While probabilistic encoding has been used in EWF rateless codes in \cite{Vukobratovic-Tmultimedia-09}, as well as in
\cite{Vukobratovic-MMSP-10} and \cite{Vukobratovic-PV-10},  an advantage of the proposed scheme in Fig. \ref{Diagram}
is that the different layers can be encoded and decoded separately.
In addition, in  \cite{Rahnavard-TIT-07} and \cite{Vukobratovic-Tmultimedia-09},  degree distributions and selection probabilities need to be optimized jointly, which is a complicated task. A practical advantage of Fig. \ref{Diagram} when applied to a multicast system for users with different bandwidth constraints,  low bandwidth clients need not receive symbols generated from source layers targeting high bandwidth clients, which reduces the complexity and time-to-decode for low BW clients.

It is worth noting that  one may alter the ordering of the output symbols from the random interleaved UEP raptor coder using scheduling algorithms while maintaining the priority of each layer. Investigations along these lines have been recently proposed in \cite{Talari-ISIT-10} and \cite{benacem-ICGCS-10}. Unlike \cite{Wei-icc-09} \cite{Wei-qbsc-10}, the proposed need not specify a packetization structure; the scheme may be  applied to data packets rather than to symbols.

\section{Problem formulations with QoS constraints}
\label{sec:problem formulation}

\subsection{Guaranteed QoS formulation}
\label{subsec:problem formulation}

We consider users that require playback media at a quality no lower than their own QoS requirement. Since the transmitter has to provide guaranteed QoS for all user classes before the start of transmission of the next source block, system delay and throughput for each source block is determined by the maximum number of transmitted symbols required to satisfy the QoS of each individual user class. As delay is a critical issue in multimedia multicast, the objective is to provide different levels of QoS guarantees according to users' requirements while minimizing total transmission overhead $\varepsilon$:

\vspace{0.1in}
\textbf{Problem} $\textbf{1.0}$ (Guaranteed QoS):
\begin{eqnarray}\label{min1}
   \min_{\rho_1,...,\rho_L} \hspace{5mm} \varepsilon
\end{eqnarray}
 \vspace{-5mm}
\begin{eqnarray}\label{suchthat1.0}
   s.t. \hspace{2mm}  \text{Prob}(PSNR_j \geq \gamma_j) \geq P_j ,   \hspace{3mm}  j=1,2,...,J,
\end{eqnarray}
where $PSNR_j$ represents the PSNR of the successfully recovered source data of the Class $j$ user given $M=(1+\varepsilon) K$ transmitted symbols, and $\gamma_j$ and $1-P_j$ denote target PSNR threshold and outage probability, respectively, for the Class $j$ user. The aim is to allocate coding rates across layers through optimization of the probabilities $\rho_l$, $1 \leq l \leq L$.

The source (e.g., video, image) coder is assumed to be progressive, so that the reconstruction media quality is determined mainly by the symbol errors in the lowest layer encountered in the recovery process. Let $q_l,l=1,2,...,L,$ represent the PSNR achieved when Layers $1$ to $l$ are successfully recovered by raptor decoding, where $q_1 \leq q_2 ... \leq q_L$. For a given source coder, if the source PSNR is represented as non-decreasing function, $f(\cdot)$, of the total number of source symbols decoded by the receiver, then $q_l=f(\sum_1^l S_l)$.  For each class $1 \leq j \leq J$, let $g_j \in \{1,2,...,L\}$ be the minimum index that satisfies $q_{g_j} \geq \gamma_j$. In order to satisfy $PSNR_{g_j} \geq \gamma_j$, users in Class $j$ require the raptor decoder to successfully decode, at minimum, Layers $1$ to $g_j$. For a given UEP raptor code design, let $P_e(l,j)$ represent the error probability that the Class $j$ decoder fails to decode layer $l$ given transmission overhead $\varepsilon$ and reception quality $\delta_j$. In the most stringent case when decoding errors across layers are independent, QoS requirements of end users can be simplified to
\begin{eqnarray}\label{QoS_Pe}
   \prod_{l=1}^{g_j}(1-P_e(l,j)) \geq P_j  \hspace{2mm} j=1,2,...,J.
\end{eqnarray}

\subsection{Best-effort QoS formulations}
\label{sec:besteffort}

While the above formulation focuses on minimizing transmission overhead subject to satisfying guaranteed user QoS, this subsection considers transmission overhead that is upper bounded due to delay constraints or cost. For this scenario, given a maximum transmission overhead, $\varepsilon_{\max}$, the service provider attempts to provide users of different classes with the best possible QoS. 
The following best-effort QoS problem extends that in \cite{Vukobratovic-Tmultimedia-09} by 1) considering both constrained and unconstrained cases, 2) allowing for allocating different weighting factors to different user classes as well as 3) possessing the previously mentioned advantages of the proposed random interleaved UEP raptor codes.

The expected PSNR of users in Class $j$, which serves as a measure of the best-effort QoS, can be evaluated as $\text{E}(PSNR_j)=\sum_{l=1}^L p_{l,j} q_l$, where $q_l$ is the PSNR achieved when Layers $1$ to $l$ are successfully recovered,  where $p_{l,j}$ represents the probability that a Class $j$ user successfully recovers Layers $1$ to $l$ but fails to recover Layer $l+1$. The optimization balances users of different classes with different channel qualities by assigning weighting coefficient $w_j$ for Class $j$ where $0 \leq w_j \leq 1$, $\sum_{j=1}^J w_j =1$. The choice of $w_j$ depends on both the user class importance as well as the number of users in that class. The weighted average PSNR over all user classes then becomes  the objective function:

\vspace{0.1in}
\textbf{Problem 2}: (Best-effort QoS)
\vspace{-2mm}\begin{eqnarray}\label{P2.0}
   \max_{\rho_1,\rho_2,...,\rho_L} \hspace{2mm}  \sum_{j=1}^J w_j  \left(  \sum_{l=1}^L p_{l,j} \cdot q_l  \right)
\end{eqnarray}
 \vspace{-5mm}
\begin{eqnarray}\label{P2.0_st}
   \text{subject to} \hspace{6mm} \varepsilon \leq \varepsilon_{\max}
\end{eqnarray}

\noindent
where 
\begin{align} \label{plj}
p_{l,j}=
\begin{cases}
 P_e(l+1, j)  \prod_{i=1}^l (1-P_e(i, j))    \hspace{4mm} &l=1,2,...,L-1 \\
   \prod_{i=1}^L (1-P_e(i, j))   \hspace{4mm} &l=L.
\end{cases}
\end{align}

In Problem $2$, no guaranteed minimum QoS is provided. For a given maximum transmission overhead, the service provider may instead aim to provide best-effort QoS to multiple user classes, but under the additional constraint of a minimum QoS guarantee for each user class:

\textbf{Problem 3}: (Best-effort QoS with constraints on individual classes)

\vspace{-2mm}\begin{eqnarray}\label{P3.0}
   \max_{\rho_1,\rho_2,...,\rho_L} \hspace{2mm}  \sum_{j=1}^J w_j \left(  \sum_{l=1}^L p_{l,j} \cdot q_l  \right)
\end{eqnarray}
 \vspace{-5mm}
\begin{eqnarray}\label{P3.0_st}
   \text{subject to} \hspace{6mm} \varepsilon \leq \varepsilon_{\max}
\end{eqnarray}
\begin{eqnarray}\label{P3.0-constraint}
    \text{and}  \hspace{2mm}  \prod_{l=1}^{g_j}(1-P_e(l, j)) \geq P_j  \hspace{2mm} j=1,2,...,J
\end{eqnarray}
where $p_{l,j}$ is given by (\ref{plj}) and $g_j \in \{ 1,2,...,L \} \hspace{2mm} (j=1,2,...J)$ is the minimum layer index that satisfies $q_{g_j} \geq \gamma_j$. Problem 2 is a special case of Problem 3 without user QoS constraints.

In the next section, we show that Problem 1.0 can be transformed to an equivalent convex optimization problem when standardized raptor codes are employed. Unfortunately, while Problems 2 and 3 cannot be similarly transformed due to the form of the $p_{l,j}$ expressions, they can still be solved numerically by searching the $(L-1)$-dimensional parameter space of $\{ \rho_1, \rho_2,..., \rho_{L-1} \}$, checking the constraints (\ref{P3.0-constraint}) and the resulting average PSNR (\ref{P3.0}). When $L=2$, the numerical method is significantly simplified as only $\rho_1 \in [0, 1]$ that gives the maximum average PSNR needs to be determined. Numerical results and comparisons for Problems 2 and 3 are provided later.

\section{Solving the guaranteed QoS problem}
\label{sec:solution}

\subsection{Evaluation of decoding failure probability}
\label{subsec:errorrate}

In the proposed design, existing high-performance standardized raptor codes can be directly applied, which enable low encoding/decoding complexity and overhead. Details about the pre-code, degree distribution and code construction can be found in \cite{3GPP-09}, (Annex B). When standardized raptor codes are employed with maximum likelihood (ML) decoding for code dimension greater than $200$, the decoding failure probability, i.e., failure to decode $k$ source symbols after $m$ symbols are successfully received, have been shown, through extensive experimentation, to be accurately modeled by \cite{luby-Tbroadcast-07},
\begin{align}\label{failure prob}
   P_{e}^r(m, k)=
   \begin{cases}
    1    \hspace{17mm}  &\mbox{if} \hspace{2mm} m \leq k \\
    a b^{m-k}  \hspace{3mm}  &\mbox{if} \hspace{2mm}  m  > k \\
\end{cases}
\end{align}
where constants $a=0.85$, $b=0.567$. For $k<200$, Eq. (\ref{failure prob}) underestimates the error probability due to short block length. One way to improve code performance for layers with fewer symbols is to merge source layers with similar optimized selection probabilities $\rho_j$  into larger layers. However, for video the condition $k < 200$ is unlikely to occur.

We also remark that standardized raptor codes outperform the recently proposed SW-raptor codes \cite{Cataldi-TIP-10}. For example, according to Fig. 2(b) of \cite{Cataldi-TIP-10}, the SW-raptor codes have a decoding failure probability of almost $100 \%$ with code dimension $K=5000$ and overhead $\varepsilon=0.03$ while standardized raptor codes have negligible decoding failure probability at the same code dimension and lower overhead $\varepsilon=0.01$ according to (\ref{failure prob}).

When more general LT or raptor codes using iterative decoding are employed, the decoding failure probability $P_e(l,j)$ can be approximated by assuming that symbol errors in iterative decoding are mutually independent
\begin{eqnarray}\label{new-layer-error}
  P_e(l,j)=1-(1-e_{l,j})^{S_l},
\end{eqnarray}
where $e_{l,j}$ is the symbol error probability of a Class $j$ user decoding Layer $l$ (also see (3) of \cite{Vukobratovic-Tmultimedia-09}) which can be analytically determined by  {\em and-or} tree analysis \cite{luby-SODA-98}. Since each layer is encoded by a separate rateless code,
evaluating the symbol error probability of each layer can be consider as a special case of (6) and (7) in
\cite{Rahnavard-TIT-07} where uniform selection is used ($k_M=1$ in \cite{Rahnavard-TIT-07}), and  Eq. (\ref{new-layer-error}) can be approximated using
\begin{align}\label{iter1}
   e_{l,j}^n=
 \begin{cases}
    1  &n=0   \\
   \exp(- \frac{t_l \delta_j}{S_l} \Omega'(1-e_{l,j}^{n-1}))  \hspace{2mm} &n \geq 1
  \end{cases}
 \end{align}
where $\Omega(.)$ is the LT code degree distribution, $\Omega'(x)$ denotes derivative with respect to $x$, $n$ is the number of decoding iterations and $t_l$ is the total number of encoded symbols transmitted for Layer $l$ in each transmission block. The asymptotic symbol error probability  $e_{l,j}=\lim_{n \rightarrow \infty} e_{l,j}^n$ of iterative decoding can be estimated by choosing a large value $n$ in Eq. (\ref{iter1}) (see \cite{luby-SODA-98}).

\subsection{Convexity analysis}
\label{sec:convex}

For a given transmission overhead $\varepsilon$, $t_l = (1+\varepsilon) K \rho_l$, \footnote {Strictly speaking, $t_l$ is a Binomial-distributed random variable with mean $(1+\varepsilon) K \rho_l$. However, the randomization of $t_l$ has little effect on the problem of interest when averaged over a large number of realizations. In addition, one can always schedule the selection of layers to make sure that $t_l$ is proportional to $\rho_l$.}
and satisfies $\sum_{l=1}^L t_l=(1+\varepsilon) K$. When standardized raptor codes are used, substituting $m=t_l \delta_j$ and $k=S_l$ into Eqs. (\ref{failure prob}) and (\ref{QoS_Pe}), and taking the logarithm of the constraints described by  (\ref{QoS_Pe}), Problem $1.0$ is transformed to:

\vspace{0.1in}
\textbf{Problem} $\textbf{1.1}$:
\begin{eqnarray}\label{min11}
   \min_{t_1,...,t_L} \hspace{10mm} \Sigma_{i=1}^L t_i
\end{eqnarray}
\begin{eqnarray}\label{suchthat}
   s.t. - \Sigma_{l=1}^{g_j} \log [1-c_l \alpha_j^{t_l} ] + \log P_j  \leq 0,   \hspace{3mm}  j=1,2,...,J,
\end{eqnarray}
where $c_l = a b^{-S_l}$, $\alpha_j = b^{\delta_j}$ and $g_j \in \{1, 2,..., L \}$. The constraint that $t_l$ is non-negative is implicitly guaranteed by the $\log(.)$ function. To ensure an integer solution, we compute $t_1, t_2,...,t_L$ as if real-valued, then round to the nearest larger integer. Although the above transformation uses the decoding failure probability evaluation of standardized raptor codes given by Eq. (\ref{failure prob}), a similar method can be applied to other decoding failure probability models that can be approximated by an exponential function.

To solve Problem $1.1$, we first prove convexity. As the objective function is linear, we only need to prove that the constraint functions are convex with respect to $t_l, l=1,2,...,L$. It can be shown that for $l=1,2,...,L,$ the second derivatives of $-\log (1-c_l \alpha_j^{t_l}) $ with respect to $t_l$ satisfy
\begin{eqnarray}\label{secondorderconditon}
   \frac{\partial^2 [-\log (1-c_l \alpha_j^{t_l})]}{ \partial t_l^2}=\frac{c_l \alpha_j^{t_l} (\log \alpha_j)^2}{(1-c_l \alpha_j^{t_l})^2 } > 0   \hspace{3mm}  j=1,2,...,J.
\end{eqnarray}
According to the second order condition of convex functions \cite{boyd-convex-04}, $- \log [1-c_l \alpha_j^{t_l}]$ is a convex function of $t_l$. Since nonnegative weighted sums preserve convexity \cite{boyd-convex-04}, the constraint functions (\ref{suchthat}) are convex functions of the vector $\mathbf{t}=[t_1, t_2,...,t_L ]^T$. Problem 1.1 can therefore be solved numerically by available convex optimization algorithms \cite{boyd-convex-04}. We remark that the above convexity holds not just for values of $a$ and $b$ in the exponential model of Eq.~(\ref{failure prob}) from \cite{luby-Tbroadcast-07} but also more generally over the range $0<a<1$ and $0<b<1$ which represent a wide family of exponential fountain code failure probability models.

Let $\mathbf{t}=[t_1,t_2,...,t_L]^T$ and $\mathbf{\lambda}=[\lambda_1,\lambda_2,...,\lambda_J]^T$ be the variable vectors of the primal and dual problems of Problem $1.1$, respectively. If $\mathbf{t^*}=[t_1^*,t_2^*,...,t_L^*]^T$ and $\mathbf{\lambda^*}=[\lambda_1^*,\lambda_2^*,...,\lambda_J^*]^T$ represent sets of primal and dual optimal points, they must satisfy the Karush-Kuhn-Tucker (KKT) optimality conditions for the objective function $f_0(.)$ and constraint functions $f_j(.)$:
\begin{eqnarray}\label{KKT1}
  f_j(\mathbf{t^*}) \leq 0, \hspace{2mm} j=1,2,...,J
\end{eqnarray}
\begin{eqnarray}\label{KKT2}
   \lambda_j^* \geq 0, \hspace{2mm}  j=1,2,...,J
\end{eqnarray}
\begin{eqnarray}\label{KKT3}
  \lambda_j^* f_j(\mathbf{t^*}) = 0, \hspace{2mm}  j=1,2,...,J
\end{eqnarray}
\begin{eqnarray}\label{KKT4}
  \nabla f_0 (\mathbf{t^*}) + \sum_{j=1}^J \lambda_j^* \nabla f_j(\mathbf{t^*}) = 0
\end{eqnarray}
where here $f_0 (\mathbf{t})=\Sigma_{i=1}^L t_i$ and $f_j(\mathbf{t})= - \Sigma_{l=1}^{g_j} \log [1-c_l \alpha_j^{t_l} ] + \log P_j, j=1,2,...,J$. Since the original Problem $1.1$ is convex and satisfies Slater's condition, the above KKT optimality conditions provide the necessary and sufficient conditions for optimality \cite{boyd-convex-04}. In general, solving the KKT condition is not straightforward. However, if we can identify a set of inequality constraints that are most likely to be active, i.e., achieve equality at the optimal solution, then we can obtain a corresponding set of primal and dual solution points and verify the optimality with KKT condition.

A simplification to Problem 1.1 arises if we have a one-to-one mapping between user classes and channel coding layers, i.e., $g_j=j$ for $j=1,2,..,J$ and $L=J$, which is the assumption used in the formulation of \cite{Vukobratovic-Tmultimedia-09}, and if all the inequality constraints are active. Using the above assumption, the solution to Problem $1.1$ can be obtained by finding $t_1$ using the constraint for Class $1$ in Eq. (\ref{suchthat}) and substituting the solution of $t_1$ into the next constraint, solving for $t_2$ with the constraint for Class $2$ in Eq. (\ref{suchthat}) etc. until all of the variables  $t_1, t_2,..., t_L$ are determined. However, since this simplification has not been proven to be equivalent to Problem 1.1 in general, the solution obtained in this manner has to be verified using the KKT optimality conditions. If all the inequality constraints are active, Eqs. (\ref{KKT1}) and (\ref{KKT3}) are automatically satisfied. Therefore, if we obtain a solution $\mathbf{t^*}$ of Problem $1.1$ by solving $f_j(\mathbf{t^*})=0, j=1,2,...,J$, we can substitute the value of $\mathbf{t^*}$ into Eq. (\ref{KKT4}) and obtain $\mathbf{\lambda^*}$. If $\mathbf{\lambda^*}$ satisfies Eq. (\ref{KKT2}), i.e.,  $\lambda_j^* \geq 0, j=1,2,...,J$, then we have proven that the value of $\mathbf{t^*}$ we obtained is indeed an optimal solution of Problem $1.1$. If the KKT optimality condition is not satisfied, then numerical methods can still be used to solve this  convex optimization problem.

\subsection{Class-to-layer mapping algorithm}
\label{sec:mapping}

In the following, we propose an algorithm to transform a general guaranteed QoS problem into a problem with one-to-one mapping between user classes and channel coding layers. The idea is to reduce the dimensionality of the problem by removing redundant user constraints and merging source-coding layers. The process is explained in the following algorithm:

\vspace{0.1in}
\textbf{Algorithm 1}: (Class-to-layer mapping algorithm)

\textbf{Step 1} (User class amalgamation): Repeat the following class amalgamation operation until $g_i < g_k$ for every $i<k$, where $1 \leq i \leq J, 1\leq k \leq J$: for any pair of user class indices $i$ and $k$ where $i<k$ (hence $\delta_i < \delta_k$), if Class $i$ users have the same or higher target PSNR threshold than Class $k$ users (i.e., $\gamma_i \geq \gamma_k$ or $g_i \geq g_k$), we absorb Class $k$ into Class $i$.

\textbf{Step 2} (Source layer merging): Repeat until for every layer $1 \leq l \leq L$, there exists a class $j, 1 \leq j \leq J$ such that $g_j=l$: if there exists a source layer $l$ where there is no corresponding user class (i.e., no $j$ exists such that $g_j=l$), Layers $l$ and $l+1$ are merged to form a new source layer $l'$ with code dimension $S_{l'}=S_{l}+S_{l+1}$.

Step 1 finds a set of the most demanding user classes with respect to their channel conditions; Step 2 reduces the number of channel coding layers to the minimum without compromising the performance. After performing Algorithm 1, we can show the following fact:

\textbf{Lemma 1}: After performing Algorithm 1, $L=J$ and $g_j=j$ for $j=1,2,...,J$. If for every Class $k$ that has been absorbed into Class $i$ in Step 1, $P_i \geq P_k$ is also satisfied, then the new optimization problem after performing Algorithm 1 is equivalent to Problem 1.1. In addition, any further partitioning of layers cannot reduce the minimum transmission overhead required to achieve the QoS requirements.

\textbf{Proof}: First we show that any QoS constraint dropped from Step 1 (user class amalgamation) is irrelevant. Suppose the QoS constraint of Class $i$ users is satisfied, i.e., $\prod_{l=1}^{g_i}(1-P_e(l,i)) \geq P_i$. Since $i<k$, we have $\delta_i<\delta_k$. Hence, Class $k$ users receive more coded symbols than Class $i$ users. Therefore, the decoding failure probability $P_e(l,i)> P_e(l,k)$ for all $1 \leq l \leq L$. Then, because $g_i \geq g_k$, from the assumption of Lemma 1, $P_i \geq P_k$, and
\begin{eqnarray}\label{QoS requirement}
   \prod_{l=1}^{g_k}(1-P_e(l,k)) & > & \prod_{l=1}^{g_k}(1-P_e(l,i)) \nonumber \\
   & \geq &  \prod_{l=1}^{g_i}(1-P_e(l,i)) \geq P_i \geq P_k.
\end{eqnarray}
Hence, the QoS constraint for Class $k$ users is also satisfied.

Next we show that after performing Algorithm 1, the number of source layers $L$ and the number of user classes $J$ are equal. The class amalgamation procedure ensures that the set $g_j$, $j=1,2,...,J$ is monotonically increasing with $j$. This fact does not change after performing the source layer merging procedure. Since $g_j \in \{1,2,...,L \}$, we have $L \geq J$. On the other hand, source layer merging ensures that for any $l=1,2,...L$, there exists an integer $j \in \{1,2,...,J \}$ such that $g_j=l$. Therefore, we also have $L \leq J$. Thus, $L=J$. Together with the fact that $g_j$ is monotonically increasing with $j$, we can conclude that $g_j=j$ for $j=1,2,...,J$.

Finally, to complete the proof, in the appendix we show that any further partitioning of layers cannot reduce the required minimum transmission overhead. \textbf{QED}.



\textbf{Remark 1}: The condition that for every Class $k$ that has been absorbed into Class $i$, $P_i \geq P_k$, is a sufficient condition for Lemma 1 but not a necessary condition. Even if this condition is not satisfied, it is possible that the transformed problem due to Algorithm 1 results in the optimal solution. In addition, if this condition is violated, to ensure that the optimal solution of the transformed problem is the optimal solution of the original problem, we can always verify if the obtained solution satisfies all the constraints of the user classes that have been amalgamated in Step 1. If not, the convex Problem 1.1 can be solved numerically. This is further illustrated in Section \ref{sec:example}.

\textbf{Remark 2}:  For best-effort QoS Problem 3, the transformation given in Algorithm 1 may not apply, as an optimal solution also depends on the fidelity measure of the multimedia source.

\textbf{Remark 3}: In the original general problem $L$ and $J$ are arbitrary, which means it is possible that a user with worse channel quality may have a higher QoS requirement. Lemma 1 and the mapping algorithm transform the original problem to a progressive transmission problem where there is a one-to-one mapping between user classes and channel coding layers.

\textbf{ Remark 4:} In the case of $L=J$,  the transmission overhead can be lower bounded by Eq.~(\ref{min OH})
which is independent of code optimization. Minimizing $\varepsilon$, as in Problems 1 and 1.1,  maximizes code performance.

\subsection{Video multicasting numerical example}
\label{sec:example}

We now illustrate the mapping process and solution to the guaranteed QoS problem for multicasting a H.264 SVC \cite{schwarz-Tcircuits-2007} video-coded stream which contains 15 layers: a base layer (BL) and 14 enhancement layers (ELs). Since our focus is on optimizing a channel coder for a given source coder, the number of information symbols and the corresponding PSNR values are taken from Table I of \cite{Vukobratovic-Tmultimedia-09}. As in \cite{Vukobratovic-Tmultimedia-09}, each source symbol represents 400 source bits. The UEP rateless encoders and decoders operate at the symbol level. We assume there are four classes of users with reception capabilities and QoS requirements shown in Table \ref{Table_1}.
\begin{table} [h] 
\caption{Example of user classes and their QoS requirements}  \label{Table_1}
{ \small 
\begin{tabular} {|l|c|c|c|c|}
\hline
User class index (j)                     &   1   &  2   &  3     &   4 \\
\hline
User reception capability $\delta_j$     &  0.4  & 0.5  & 0.6    &   1  \\
\hline
User QoS req. (PSNR thr. $\gamma_j$(dB))        & 25.79 & 29   & 27.25  & 40.28 \\
\hline
\# Decoded symbols to achieve QoS    & 400 & 1155   & 700  & 3800 \\
\hline
\# Decoded source layers required ($g_j$)      &  1    & 4    &  2     &  15 \\
\hline
Probability threshold $P_l$  & 0.8   & 0.9  & 0.85   & 0.95  \\
\hline
\end{tabular}}
\end{table}

Using the previously described simplification strategy for class and layer mapping, we observe that $g_2>g_3$ while $\delta_2<\delta_3$, which means that Class $3$ users have both better reception capabilities and lower PSNR requirements than Class $2$ users. Therefore, the QoS constraint from Class $3$ users can be dropped. Then, since the number of layers required by the three classes are $1$, $4$ and $15$, after the layer-merging procedure of Algorithm 1, we obtain a new set of channel layers with Layer 1 comprising the BL, Layer 2 consisting of the first 3 ELs, and Layer 3 consisting of the fourth to fourteenth ELs. Since $P_2 >P_3$, from Lemma 1, the new problem after mapping is equivalent to the original problem. The parameters of the transformed problem after the mapping are shown in Table \ref{Table_2}.
\begin{table} [h] \caption{User classes and QoS requirements after the mapping.} \label{Table_2}
{\small 
\begin{tabular} {|l|c|c|c|c|}
\hline
Combined class-layer index (j or l) &   1   &  2   &  3 \\
\hline
Reception capability $\delta_j$     &  0.4  & 0.5  &   1  \\
\hline
PSNR threshold $\gamma_j$(dB)        & 25.79 & 29   & 40.28 \\
\hline
Number of decoded symbols to achieve QoS    & 400 & 1155   & 3800 \\
\hline
Number of decoded layers required $g_j=j$   &  1    & 2   &  3   \\
\hline
Probability threshold $P_j$  & 0.8   & 0.9 & 0.95  \\
\hline
Number of symbols $S_l$ in each layer     &400    &755  & 2645  \\
\hline
\end{tabular} }
\end{table}

To determine  the interleaving probabilities for the standardized raptor codes for the three new layers, $\rho_1$, $\rho_2$ and $\rho_3$ need to be determined to minimize  $t_1+t_2+t_3$ such that
\begin{eqnarray}\label{problem}
\begin{cases}
  (1-a b^ {(t_1 \delta_1-S_1)} ) \geq P_1   \\
  (1-a b^ {(t_1  \delta_2-S_1)}) (1-a b^ {(t_2 \delta_2-S_2)}) \geq P_2  \\
  (1-a b^ {(t_1  \delta_3-S_1)}) (1-a  b^ {(t_2 \delta_3-S_2)}) (1-a b^ {(t_3  \delta_3-S_3)}) \geq P_3.
\end{cases}
\end{eqnarray}

Assuming all the inequality constraints are active, we obtain a minimum overhead $\varepsilon_{min}=36.2\% $, which is achieved when $\rho_1=0.1946$, $\rho_2=0.2933$ and $\rho_3=0.5121$. The solution is then verified to be optimal using KKT conditions. In contrast, equal error protection (EEP) allocation requires a minimum overhead of $152 \%$, a factor of over four higher.

With the optimal selection parameters, we find that Class 3 users of the original problem (Table \ref{Table_1}) can successfully decode the base layer and one enhancement layer with a probability higher than $99.9 \% $. This means that even if the target probability threshold $P_3 =99 \%$ in Table \ref{Table_1}, which violates the assumption of Lemma 1, the problem transformed by Algorithm 1 still has the same optimal solution as the original problem. As a further remark, let us suppose that the conditions of Lemma 1 were violated, and we assume the extreme case of $P_3=99 \%$ and vary the value of $\delta_3$ within the range $0.5=\delta_2 < \delta_3 \leq 1=\delta_4$. In that case, only when $0.5< \delta_3 < 0.503$, our obtained solution does not satisfy the QoS constraint of Class $3$ users. In practice,  however, distinct classes would have a greater reception capability difference than $\frac{0.003}{0.500} \times 100 \% =0.6 \%$.

\section{Numerical and simulation results}
\label{sec:comparison}

This section provides comparisons of the proposed random interleaved UEP
rateless code design to EEP codes and to other recent UEP rateless codes. The parameters of the different scenarios are described in the corresponding figure captions. Performance of LT codes are evaluated using {\em and-or} tree analysis while standardized raptor codes are evaluated
using Eq. (\ref{new-layer-error}). Simulations are also used to confirm the {\em and-or} tree analysis.

\begin{figure}
\begin{center}
\includegraphics[width=3.6in, keepaspectratio] {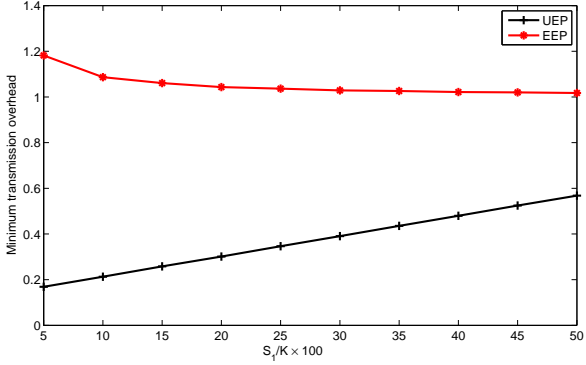}
\caption{UEP versus EEP for varying $S_1/S_2$, standardized raptor codes, $K=1155;\delta=[0.5,0.9];P=[0.95,0.9].$}
\label{UEP_EPP_cpr1}
\end{center}
\end{figure}

\begin{figure}
\begin{center}
\includegraphics[width=3.6in, keepaspectratio] {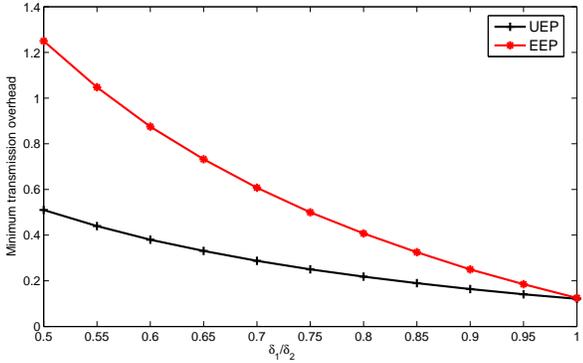}
\caption{UEP versus EEP for varying $\delta_1/\delta_2$, standardized raptor codes, $S=[400,755]; \delta_2=0.9; P=[0.95,0.9].$}
\label{UEP_EPP_cpr2}
\end{center}
\end{figure}

\begin{figure}[htb]
\begin{center}
\includegraphics[width=3.6in, keepaspectratio]
{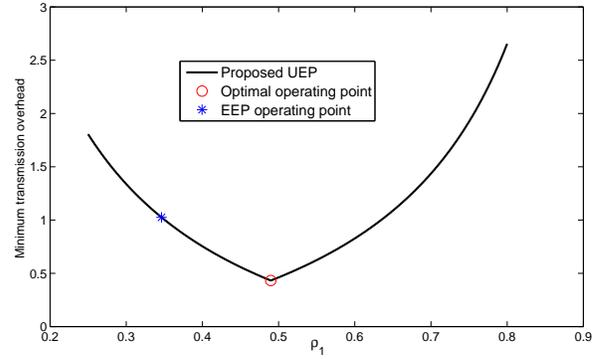} \caption{ The effect of layer allocation probability $\rho_1$, standardized raptor codes, $L=2; K=1155;S=[400,755] \delta=[0.5,0.9];P=[0.95,0.9].$} \label{UEP_rho}
\end{center}
\end{figure}

Figs. \ref{UEP_EPP_cpr1} to \ref{UEP_rho}  compare  the proposed UEP
design to EEP design for the guaranteed QoS problem when standardized raptor

codes are employed. The minimum transmission overhead is evaluated using
the method described in Section~\ref{sec:example}. For simplicity, only two layers are considered. The dimension of the standardized raptor code used in layer $l$ is $S_l$. The inefficiencies incurred by the standardized raptor codes are characterized by the decoding failure probability $P_e^r(.)$ in Eq. (\ref{failure prob}) and are small as expected. The optimal selection probability $\rho_1$ and minimum overhead for the UEP scheme are obtained by the simplified method described in Section \ref{sec:solution} for solving Problem $1.1$, i.e., by assuming that all inequality constraints are active. All results shown in Figs. \ref{UEP_EPP_cpr1} to \ref{UEP_rho} were verified to satisfy the KKT optimality conditions. To achieve EEP,
the ratio $\rho_1/\rho_2$ is fixed to $S_1/S_2$.  Fig. \ref{UEP_EPP_cpr1} shows minimum transmission overhead, $\varepsilon$, required for optimized UEP and EEP raptor codes as the ratio between the numbers of bits in the two layers
is varied. Fig. \ref{UEP_EPP_cpr2}  compares UEP and EEP as a function of  channel reception quality  of the first user class,  $\delta_1$. It can be seen that UEP has a significant advantage over EEP whenever the channel reception qualities of the two classes differ appreciably. Fig. \ref{UEP_rho} plots minimum transmission overhead, $\varepsilon$,  as a function of selection probability $\rho_1$ where it is observed that $\varepsilon$ is very sensitive to the choice of $\rho_1$. In particular, a non-optimized allocation scheme may be significantly outperformed  by EEP.

To enable comparison with the UEP raptor code from \cite{Rahnavard-TIT-07} as well as with EWF codes from \cite{Vukobratovic-Tmultimedia-09}, rather than use raptor codes, we employ iteratively decoded  LT codes that have degree distribution \cite{shokrollahi-tit-06}
\begin{eqnarray}\label{inner_LT_distribution}
\nonumber \Omega_r(x)=0.007969x+0.493570x^2+0.166622x^3
\\\nonumber +0.072646x^4+0.082558x^5+0.056058x^8+0.037229x^9
\\+0.055590x^{19}+0.025023x^{65}+0.003135x^{66}
\end{eqnarray}
for all layers as used
for UEP codes in \cite{Rahnavard-TIT-07} and for the EWF code \cite{Vukobratovic-Tmultimedia-09}. That is, for analytical simplicity, no pre-code is used in any of the schemes. The decoding failure probability $P_e$ on the left side of the constraint functions in Eq. (\ref{QoS_Pe}) is evaluated as follows: 
the symbol error probability $e_l$ of Layer $l$ for the UEP rateless codes in \cite{Rahnavard-TIT-07}, the EWF code, and the proposed random interleaved scheme are estimated by {\em and-or} tree analysis and obtained using Eqs. (6) and (7) in \cite{Rahnavard-TIT-07}, Eq. (7) in \cite{Sejdinovic-Tcom-09}, and Eq. (\ref{iter1}) in this paper, respectively. The failure probability of decoding each layer is estimated as $P_e(l)=1-(1-e_l)^{S_l}$.

Parameter optimization of the other schemes can be found in \cite{Cao-bmsb-10}  and are not reproduced here. Fig. 5 in \cite{Cao-bmsb-10} provides the minimum transmission overhead required to satisfy all the user constraints of the proposed random interleaved scheme as well as that in \cite{Rahnavard-TIT-07} using different values of $k_M$, a parameter that governs the degree of non-uniformity of input symbol selection. Fig. 6 in \cite{Cao-bmsb-10} shows a similar comparison between the proposed scheme and the EWF code. The size of the first window in the EWF code is fixed to the number of symbols in Layer $1$ ($S_1$). Parameter $\Gamma_1$ is the probability of choosing the more important first layer during encoding (see \cite{Vukobratovic-Tmultimedia-09}). It can be observed that when all schemes are optimized, the proposed random interleaved rateless code performance matches that of  \cite{Rahnavard-TIT-07} as well as \cite{Sejdinovic-Tcom-09}. The existence of two local minima in Fig. 6 in \cite{Cao-bmsb-10} is due to the symbol error rates of the more important bits not  decreasing monotonically as $\Gamma_1$ increases (see Fig. $1$ in \cite{Vukobratovic-Tmultimedia-09}).

\begin{figure}[htb]
\begin{center}
\includegraphics[width=3.6in, keepaspectratio]{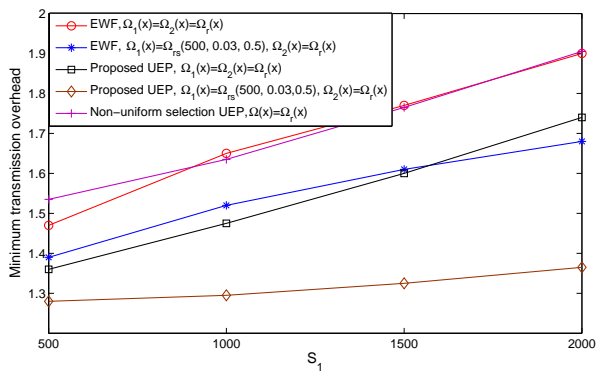}  
\caption{Performance comparisons using LT codes with different degree distributions with $L=2; K=9000; \delta=[0.4,0.8]; P=[0.95,0.8].$ Also shown are EWF from \cite{Sejdinovic-Tcom-09} and non-uniform selection  from \cite{Rahnavard-TIT-07}. } 
\label{Fig_RSD_newcp}
\end{center}
\end{figure}

An advantage of EWF codes \cite{Sejdinovic-Tcom-09} over those in \cite{Rahnavard-TIT-07} is flexibility in deploying different degree distributions for different windows.
Fig. \ref{Fig_RSD_newcp} plots transmission overhead as a function of numbers of symbols in the first layer or window for the three UEP schemes each using LT codes after optimization over their respective parameters where different degree distributions are applied to different EWF code windows as well as to different layers of the proposed UEP scheme.
Degree distributions, chosen for the more important bits (MIB) and less important bits (LIB), are denoted as $\Omega_1(x)$ and $\Omega_2(x)$, respectively.
Degree distribution $\Omega_r(x)$ described by Eq. (\ref{inner_LT_distribution}) is used as well as  a truncated robust soliton distribution (RSD) $\Omega_{rs}(k_{rs}, \delta, c)$, where $k_{rs}$ is the maximum degree, is applied to the MIB for the EWF  and  proposed random interleaved schemes. The truncated RSD has better error performance compared to $\Omega_r(x)$ at the cost of higher decoding complexity. It can be seen from Fig. \ref{Fig_RSD_newcp} that the truncated RSD for the MIB provides a significant performance boost for both schemes. When the same degree distributions are used, the proposed random interleaved scheme matches the performance of existing schemes.

We note that  in Fig. \ref{UEP_rho}, $\varepsilon_{\min}$ = 0.432, indicating that the code performance nearly achieves minimum overhead while the code performance shown in Fig. \ref{Fig_RSD_newcp} does not come close to the minimum overhead. This difference is mainly attributable to the use of standardized raptor codes, which includes a high performance pre-code as well as efficient maximum likelihood (ML) decoding in contrast to the iterative decoding used for Fig. \ref{Fig_RSD_newcp}. It can be argued that the performance of existing UEP designs in \cite{Rahnavard-TIT-07} and \cite{Sejdinovic-Tcom-09} can similarly benefit from a precode and ML decoding.  However, ML decoding complexity of the proposed random-interleaved design would likely be lower due to a lower dimension decoding matrix obtained from separate-layer decoding. In addition, the code structure and generating matrix of systematic standardized raptor code implementation has been highly optimized, including the decoding schedule in the code constraint processor  \cite{Luby-RaptorFEC-07}. To the authors' best knowledge, such techniques have not been applied to EWF codes, which may also be complicated by their overlapping structure.

\begin{figure}[htb]
\begin{center}
\includegraphics[width=3.6in, keepaspectratio]{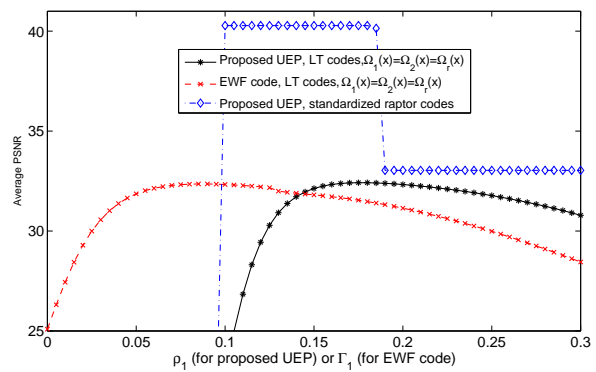}
\caption{Average PSNR performance of UEP schemes. LT codes with iterative decoding, $L=2; S=[400,3400];\delta=[0.55,1]; P=[0.95,0.8]; \varepsilon_{\max}=1; w=[0.5,0.5].$} \label{Fig_PSNRdb_cp1}
\end{center}
\end{figure}

Fig. \ref{Fig_PSNRdb_cp1} shows source reconstruction quality, in terms of  PSNR, of the proposed random interleaved and EWF schemes for the best-effort QoS formulations of Problems $2$ and $3$. Transmission of H.264 SVC coded CIF \textit{Stefan} video sequence \cite{Vukobratovic-Tmultimedia-09} is performed in two layers, with the first (base) layer containing $S_1=400$ symbols and all enhancement streams comprising the second layer with $S_2=3400$ symbols. 
Successfully decoding the first layer provides a PSNR of 25.79 dB while decoding both layers provides a PSNR of 40.28 dB. Performance is plotted
as average PSNR versus selection probability $\rho_1$ for the proposed random interleaved scheme and the first window selection probability $\Gamma_1$ of the EWF code. Given $\rho_1$ or $\Gamma_1$, average PSNR is obtained numerically by setting $\varepsilon = \varepsilon_{\max}$ and substituting the corresponding decoding failure probabilities $p_e(i,j)$ into (\ref{plj}) and (\ref{P3.0}). It should be noted that selection probabilities $\rho_1$ and $\Gamma_1$ for the two different schemes are not directly comparable.
For the cross-marked and star-marked curves, we have used the LT code with an iterative decoder and degree distribution $\Omega_r(x)$ applied to all windows and layers. For these parameters, both the proposed random interleaved and EWF schemes provide a maximum average PSNR of around 32.4 dB. For Problem 3, the feasible regions of selection probabilities $\rho_1$ and $\Gamma_1$ are obtained by checking constraints (\ref{P3.0-constraint}). We note that for Problem 3, the maximum achievable average PSNRs remain the same since both optimal operating points of the proposed UEP scheme and the EWF code lie inside the feasible regions. The diamond-marked curve shows the results when standardized raptor codes are employed for the proposed random interleaved UEP scheme. A maximum average PSNR of 40.28 dB can be achieved for $0.11 \leq \rho_1 \leq 0.18$, which, as expected, is significantly higher than the other two LT coded curves. We can also observe from Fig. \ref{Fig_PSNRdb_cp1} that different choices of $\rho_1$ result in significant differences in average PSNR, showing the need for optimization. Finally, we observe that the steep performance curve of standardized raptor codes results in only two obtained PSNR values.

\begin{figure}[htb]
\begin{center}
\includegraphics[width=3.6in, keepaspectratio]{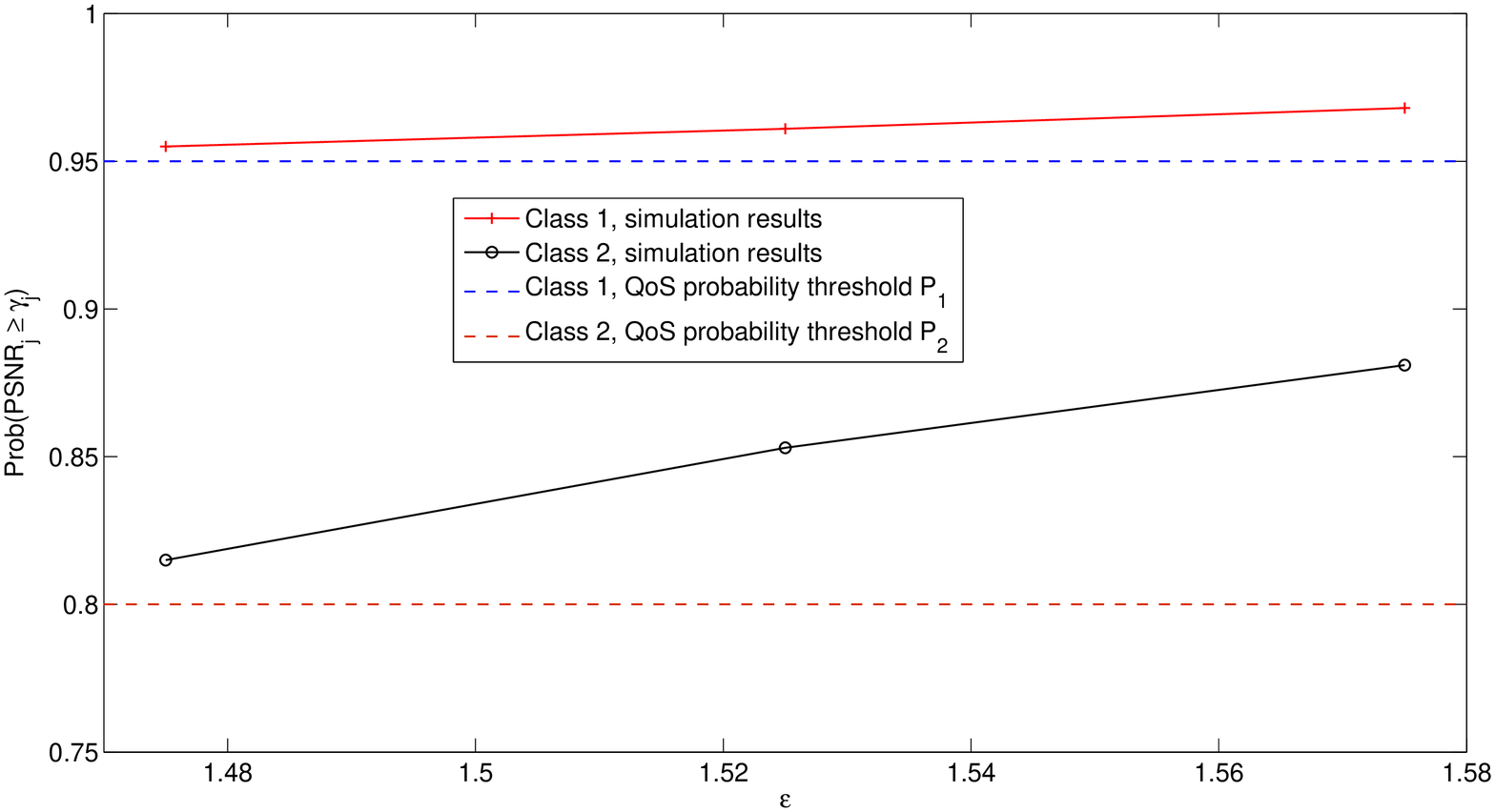} \vspace{-5mm}\caption{Outage probability comparison, simulation results versus analysis for given desired thresholds, $L=2; S=[1000,8000];\delta=[0.4,0.8]; P=[0.95,0.8]; \rho_1=0.19. $}
\label{LT_simu1}
\end{center}
\end{figure}

The above LT coding / iterative decoding results are obtained using  {\em and-or} tree analysis which assumes infinite block length. As a check,  simulation of  LT codes with degree distribution $\Omega_r(x)$ and iterative belief propagation (BP)  decoding \cite{luby-focs-02} \cite{shokrollahi-tit-06} are provided in Fig \ref{LT_simu1}.  Layer selection parameter $\rho_1=0.19$ obtained from Eqs. (\ref{new-layer-error}) and (\ref{iter1}) determine the constraints in Problem 1.0. The horizontal axis depicting transmission overhead includes the minimum overhead achieved by {\em and-or} tree analysis ($\varepsilon_{and-or}=1.475$), as well as $5 \%$ (1.525) and $10 \%$ (1.575) greater than the minimum. The resulting PSNR for each user class is computed for each realization. The vertical axis shows the relative frequency that the PSNR is larger than the desired threshold ($ \text{Prob}(PSNR_j \geq \gamma_j)$) for each of the two user classes. It can be seen from the left side of Fig. \ref{LT_simu1} that the simulation results closely match the {\em and-or} tree analysis. The probability of reaching target PSNR $\gamma_j$ in simulation is very close to the desired probability threshold $P_j$. Also, by increasing the overhead to  $1.525$, a higher probability in reaching target QoS can be obtained.

\vspace{-5mm}
\section{Conclusions}
\label{sec:conclusions}
A randomly interleaved rateless coder for scalable multimedia multicasting systems with heterogeneous users is optimized for guaranteed and best-effort QoS. The resulting design achieves unequal error protection. Further, guaranteeing QoS is shown to be a convex optimization problem, which can be solved analytically in practical scenarios. Numerical results show the transmission overhead required for the optimized proposed UEP rateless codes to be significantly less than that for EEP design and at least as low as recent optimized EWF and non-uniform-selection UEP rateless code designs.  Significant gains for the proposed UEP scheme can be obtained by employing standardized raptor codes. For example, in the best-effort QoS example in Fig. \ref{Fig_PSNRdb_cp1}, the maximum achievable average PSNR using the proposed design employing standardized raptor codes is about $8$ dB higher than that of either the proposed or EWF designs based on LT codes with iterative decoding.

\appendix
\label{appendix}
We prove the last part of Lemma 1, i.e., after performing Algorithm 1, transmission overhead cannot be further reduced with additional layer partitioning and selection probability re-assignment. Let Scheme A denote the source-to-channel layer mapping produced by Algorithm 1 and denote Scheme B as one which further partitions Layer $l$ into Layers $m$ and $n$ with dimensions $S_m$ and $S_n$, respectively. Denote the resulting optimal selection probabilities for Scheme B which minimize the transmission overhead as $\rho_m$ and $\rho_n$ for Layers $m$ and $n$, respectively. We now show that the minimum required transmission overhead is no larger by using Scheme A with selection probability $\rho_l=\rho_m+\rho_n$ assigned to Layer $l$. For the same number of total transmitted symbols $M$, the effective average raptor code rates for Layer $l$ in Scheme $A$, Layer $m$ in Scheme $B$ and Layer $n$ in Scheme $B$ are $R_l=\frac{S_l}{M \rho_l}$, $R_m=\frac{S_m}{M \rho_m}$ and $R_n=\frac{S_n}{M \rho_n}$, respectively. Without loss of generality, we assume $\rho_m/S_m \geq \rho_n/S_n$. Then it can be shown that $R_l=\frac{S_m+S_n}{M (\rho_m+\rho_n)} \leq \frac{\frac{S_n}{\rho_n} \rho_m + \frac{S_n}{\rho_n} \rho_n}{M (\rho_m+\rho_n)} = \frac{S_n}{M \rho_n} =R_n$. As the decoding failure probability of the raptor codes is monotonically increasing with code rate for the same user class, we have $ (1- P_e(l, j)) \geq (1- P_e(n,j)) > (1- P_e(m,j))(1- P_e(n,j))$ for any class index $j$, where $P_e(.)$ is the same decoding failure probability function as defined in (\ref{QoS_Pe}). This means that for the same number of transmitted symbols, the original mapping scheme (Scheme A) has higher probability of successfully decoding all the symbols in Layer $l$ than Scheme B for all user classes. Therefore, for the same QoS constraints described by (\ref{QoS_Pe}), Scheme A requires less minimum transmission overhead compared to Scheme B. Finally, raptor codes with larger dimension have better performance for the same code rate, which also implies no further layer partitioning.

\bibliographystyle{jcn}

\begin{thebibliography}{12}

\bibitem{Cao-isit-10} Y. Cao, S.D. Blostein and W.Y. Chan, ``Unequal error protection rateless coding design for multimedia multicasting", {\em Proc. IEEE Int. Symp. Information Theory}, pp. 2348-2442, June 2010.

\bibitem{Cao-bmsb-10} Y. Cao, S.D. Blostein and W.Y. Chan, ``Optimization of rateless coding for multimedia multicasting", {\em Proc. IEEE Int. Symp.  Broadband Multimedia Systems and Broadcasting}, March 2010.

\bibitem{schwarz-Tcircuits-2007} H. Schwarz, D. Marpe, and T. Wiegand, ``Overview of the scalable
video coding extension of the H.264/AVC standard,"  {\em IEEE Trans. Circuits Syst. Video Technol.}, vol. 17, no. 9, pp. 1103-1120, Sep. 2007.

\bibitem{Mohr-JSAC-00} A.E. Mohr, E.A. Riskin and R.E. Ladner, ``Unequal loss protection: graceful degradation of image quality over packet erasure channels through forward error correction", {\em IEEE JSAC}, vol. 18, no. 6, pp. 819 - 828, 2000.

\bibitem{Raja-EJASP-07} N. Raja, Z. Xiong, and M. Fossorier, ``Combined source-channel coding of images under power and bandwidth constraints," {\em EURASIP J. Advances Signal Processing}, 2007.

\bibitem{Jaspar-ProcIEEE-07}  X. Jaspar, C. Guillemot, and L. Vandendorpe, ``Joint source-channel turbo techniques for discrete-valued sources: From theory to practice," {\em Proc. IEEE}, vol. 95, no. 6, pp. 1345-1361, Jun. 2007.

\bibitem{Xu-JSAC-07} Q. Xu, V. Stankovic, and Z. Xiong, ``Distributed joint source-channel coding of video using Raptor codes," {\em  IEEE J. Select. Areas Commun.}, vol. 25, no. 4, pp. 851-861, May 2007.

\bibitem{Ahmad-TCSVC-10} S. Ahmad, R. Hamzaoui, and M. Al-Akaidi, ``Adaptive unicast video streaming with rateless codes and feedback," {\em IEEE Trans. Circuits Syst. Video Technol.}, vol. 20, no. 2, pp. 275-285, Feb. 2010.

\bibitem{Liu-JSAC-10} Z. Liu, Z. Wu, P. Liu, H. Liu, and Y. Wang, ``Layer bargaining: Multicast layered video over wireless networks," {\em  IEEE J. Select. Areas Commun.}, vol. 28, no. 3, pp. 445-455, Apr. 2010.

\bibitem{VdS-TMC-06}  M. Van der Schaar, Y. Andreopoulos,  Z. Hu, ``Optimized scalable video streaming over IEEE 802.11a/e HCCA wireless networks under delay constraints," {\em IEEE Trans. Mobile Comput.}, vol. 5, no. 6, pp. 755-768, 2006.

\bibitem{Maani-TSVT-10} E. Maani and A. K. Katsaggelos, ``Unequal error protection for robust streaming of scalable video over packet lossy networks," {\em  IEEE Trans. Circuits Syst. Video Technol.}, vol. 20, no. 3, pp. 407-416, Mar. 2010.

\bibitem{Lin-04} S. Lin and D.J. Costello, ``Error Control Coding", Amazon, 2004.

\bibitem{mitzenmacher-itworkshop-04} M. Mitzenmacher, ``Digital fountains: A survey and look forward," {\em Proc. IEEE Info. Theory Workshop.}, San Antonio, pp. 271-276, Oct. 2004.

\bibitem{shokrollahi-tit-06} A. Shokrollahi, ``Raptor Codes," {\em IEEE Trans. Info. Theory}, vol. 52, no. 6, pp. 2551-2567, Jun. 2006.

\bibitem{3GPP-09}  {\em 3rd Generation Partnership Project;Technical Specification Group Services and System Aspects; Multimedia Broadcast/Multicast Service (MBMS);Protocols and codecs, 3GPP TS 26.346 V8.3.0}, Jun. 2009.

\bibitem{luby-Tbroadcast-07} M. Luby, T. Gasiba, T. Stockhammer, and M. Watson, ``Reliable multimedia download delivery in cellular broadcast networks," {\em IEEE Trans.  Broadcasting}, vol. 53, no. 1, pp. 235-246, Mar. 2007.

\bibitem{Luby-RaptorQ-11} M. Luby and T. Stockhammer, ``Universal object delivery using RaptorQ", IETF RMT Working Group, Work in Progress: "draft-luby-uod-raptorq-00", 2011. (http://tools.ietf.org/html/draft-luby-uod-raptorq-01)

\bibitem{Luby-09} M. Luby, ``Raptor codes," {\em Foundations and Trends in Communications and Information Theory,} vol. 6, no. 3-4, pp. 213-322, 2009.

\bibitem{Schierl-WCM-06} T. Schierl, K. Gaenger, C. Hellge, T. Stockhammer, T. Wiegand, ``SVC-based multisource streaming for robust video transmission in mobile ad-hoc networks," {\em IEEE Wireless Comm. Mag.}, v. 13, no. 5, pp. 96-103, 2006.

\bibitem{Wagner-ICME-06}  J.-P. Wagner, J. Chakareski, P. Frossard, ``Streaming of scalable video from multiple servers using rateless codes," {\em Proc. ICME}, pp. 1501-1504, 2006.

\bibitem{Ji-TMM-12} W. Ji, Z. Li and Y. Chen, ``Joint source-channel coding and optimization for layered video broadcasting to heterogeneous devices", {\em IEEE Trans.  Multimedia}, vol. 14, no. 2, pp. 443-455, April 2012.

\bibitem{Xu-TCSVT-07} Q. Xu, V. Stankovic and Z. Xiong, ``Wyner-Ziv video compression and fountain codes for receiver-driven layered multicast", {\em IEEE Trans. Circuits and Systems for Video Technol.}, vol. 17. no. 7, July 2007.

\bibitem{Golrezaei-Magzine-13} N.Golrezaei, A.F. Molisch, A.G. Dimakis and G. Caire, ``Femtocaching and device-to-device collaboration: a new architecture for wireless video distribution", {\em IEEE Commun. Mag.}, vol. 51, no. 4, pp. 142-149, 2013.

\bibitem{Shanmugam-TIT-13} K. Shanmugam, N. Golrezaei, A.F. Molisch, A.G. Dimakis and G. Caire, ``Femtocaching: wireless content delivery through distributed caching helpers", {\em IEEE T. Info. Theory}, vol. 59, no. 12, pp. 8402-8413, 2013.

\bibitem{Albanese-Tit-96}  A. Albanese et. al., ``Priority encoding transmission", {\em IEEE Trans. Info. Theory}, vol. 42, no. 6, pp. 1737-1744,  Nov. 1996.

\bibitem{Chou-pvw-03} P.A. Chou, H.J. Wang and V.N. Padmannabhan, ``Layered multiple description coding", {\em Proc. Packet Video Workshop}, Nantes, 2003.

\bibitem{Hamzaoui-SPM-05}  R. Hamzaoui, V. Stankovic, Z. Xiong, ``Optimized error protection of scalable image bitstreams," {\em IEEE Signal Processing Mag.}, vol. 22, no. 6, pp. 91-107, Nov. 2005.

\bibitem{Stankovic-SP-05} V. Stankovic, R. Hamzaoui, and Z. Xiong, ``Robust layered multiple description coding of scalable media data for multicast," {\em IEEE Signal Processing Letters}, vol. 12, no. 2, pp. 154-157, Feb. 2005.

\bibitem{Chou-TMultimedia-01} P.A. Chou, A.E. Mohr, A. Wang, and S. Mehrotra, ``Error control for receiver-driven layered multicast of audio and video", {\em  IEEE Trans. Multimedia}, vol. 3, no. 1, pp. 108-122,  2001.

\bibitem{Chou-TMultimedia-06} P.A. Chou and Z. Miao, ``Rate-distortion optimized streaming of packetized media", {\em  IEEE Trans. Multimedia}, vol. 8, no. 2, pp. 390-404, 2006.

\bibitem{Rahnavard-TIT-07}  N. Rahnavard, B.N. Vellambi, and F. Fekri, ``Rateless codes with unequal error protection property," {\em IEEE Trans. Info. Theory},  vol. 53, no. 4, pp. 1521-1532, Apr. 2007.

\bibitem{Talari-Milcom-09} A. Talari and N. Rahnavard, ``Unequal error protection rateless coding for efficient MPEG video transmission," {\em Proc. MILCOM}, pp. 1-7, 2009.

\bibitem{Sejdinovic-Tcom-09} D. Sejdinovic, D. Vukobratovic, A. Doufexi, V. Senk and R. Piechocki, ``Expanding window fountain codes for unequal error protection," {\em IEEE Trans. on Communications}, vol. 57, no. 9, pp. 2510-2516, Sep. 2009.

\bibitem{Vukobratovic-Tmultimedia-09} D. Vukobratovic, V. Stankovic, D. Sejdinovic, L. Stankovic, and Z. Xiong, ``Scalable Video Multicast Using Expanding Window Fountain Codes,"  {\em IEEE Trans. Multimedia}, vol. 11, no. 6, pp. 1094-1104, Oct. 2009.

\bibitem{Studlholme-ISIT-06} C. Studlholme and I. Blake, ``Windowed erasure codes," in {\em Proc. Intl. Symp. Inform. Theory (ISIT)}, pp. 509-513, July 2006.

\bibitem{Bogino-ISCAS-07} M.C.O. Bogino, P. Cataldi, M. Grangetto, E. Magli, and G. Olmo, ``Sliding-window digital fountain codes for streaming multimedia contents," in {\em Proc. IEEE ISCAS}, pp. 3467-3470, 2007.

\bibitem{Cataldi-TIP-10} P. Cataldi, M. Grangetto, T. Tillo, E. Magli, and G. Olmo, ``Sliding-window Raptor codes for efficient scalable wireless video broadcasting with unequal loss protection," {\em IEEE Trans. on Image Processing}, vol. 19, no. 6, pp. 1491-1503, June 2010.

\bibitem{Ahmad-TMM-11} S. Ahmad, R. Hamzaoui, and M. M. Al-Akaidi, ``Unequal error protection
using fountain codes with applications to video communication," {\em IEEE Trans. Multimedia}, vol. 13, no. 1, pp. 92-101, Feb. 2011.

\bibitem{luby-focs-02} M. Luby, ``LT codes," {\em Proc. 43rd Annu. IEEE Symp. Foundations of Computer Science}, Vancouver, Canada, pp. 271-282, Nov. 2002.

\bibitem{Yang-Tcomputers-12} K. Yang and J. Wang, ``Unequal error protection for streaming media based on rateless codes", {\em IEEE Tran. Comput.}, vol. 61, no. 5, pp. 666-675, 2012.

\bibitem{Vukobratovic-MMSP-10} D. Vukobratovic and V. Stankovic, ``Unequal error protection random linear coding for multimedia communications", {\em Proc. IEEE International Workshop
on Multimedia Signal Processing}, pp. 280-285, Oct. 2010.

\bibitem{Vukobratovic-PV-10} D. Vukobratovic and V. Stankovic, ``On unequal error protection random linear coding for scalable video broadcasting", {\em Proc. Packet video workshop}, pp. 48-55, Dec. 2010.

\bibitem{Talari-ISIT-10} A. Talari, B. Hahrasbi, N. Rahnavard, ``Efficient symbol sorting for high intermediate recovery rate of LT codes", {\em  Proc. ISIT}, pp. 2443-2447,  2010.

\bibitem{benacem-ICGCS-10} L. Benacem and S.D. Blostein, ``Raptor-network coding enabled strategies for energy saving in DVB-H  multimedia communications", {\em Proc. First Int. Conf. on Green Circuits and Systems}, pp. 527-532, Jun. 2010.

\bibitem{Wei-icc-09} W. Sheng, W.Y. Chan, S.D. Blostein and Y. Cao, ``Asynchronous and Reliable Multimedia Multicast with Heterogeneous QoS Constraints", {\em Proc. IEEE ICC}, May 2010.

\bibitem{Wei-qbsc-10} W. Sheng, W-.Y. Chan, S.D. Blostein, ``Rateless code based multimedia multicasting with outage probability constraints", {\em 25th Biennial Symposium on Communications}, pp. 134-138, May 2010.

\bibitem{luby-SODA-98} M. Luby, M. Mitzenmacher, and A. Shokrollahi, ``Analysis of random processes via and-or tree evaluation," {\em in Proc. 9th SIAM Symp. Discrete Algorithms (SODA)}, pp. 364-373, Jan. 1998.

\bibitem{boyd-convex-04} S.P. Boyd and L. Vandenberghe, ``Convex Optimization", {\em Cambridge University Press}, 2004.


\bibitem{Luby-RaptorFEC-07} M. Luby, A. Shokrollahi, M. Watson and T. Stockhammer, ``Raptor forward error correction scheme for object delivery", IETF RFC 5053, Oct 2007, available at ``http://tools.ietf.org/html/rfc5053".


\end{thebibliography}

\epsfysize=3.2cm
\begin{biography}{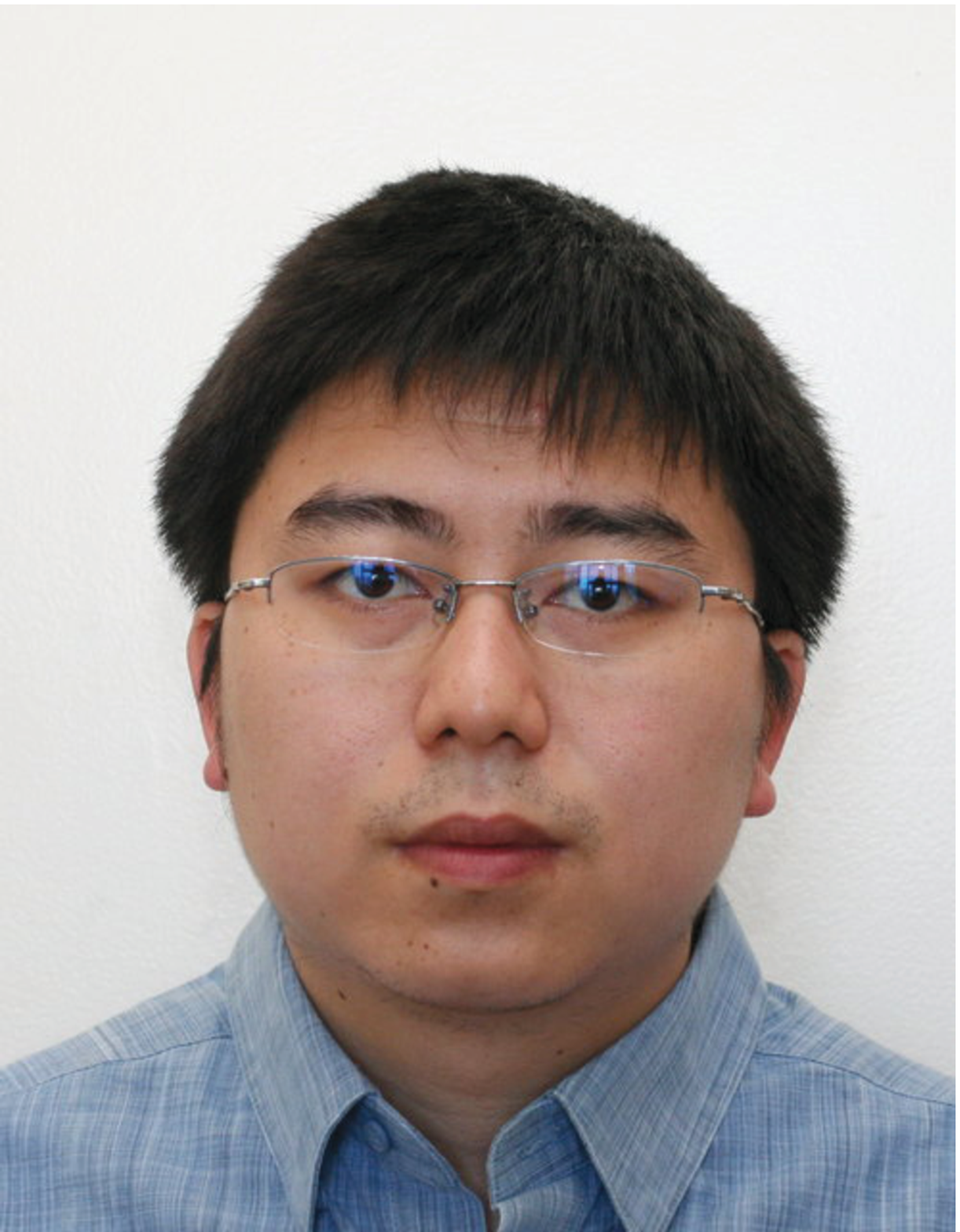}{Yu Cao}
received his B.S. degree from Department of Electronic Engineering, Tsinghua University, Beijing, China and his M.S. and Ph.D. degrees from Department of Electrical and Computer Engineering, Queen's University, Kingston, Canada. From 2011 to 2012, he was a post-doctoral fellow at Queen's University, Canada. Since 2012, he has been working as a research engineer at Huawei Canada Research Center, Ottawa, Ontario, Canada. His research interests are forward error correction codes and optimization for wireless communications. His current research focus is air interface design for next generation radio access networks. 
\end{biography}

\epsfysize=3.2cm
\begin{biography}{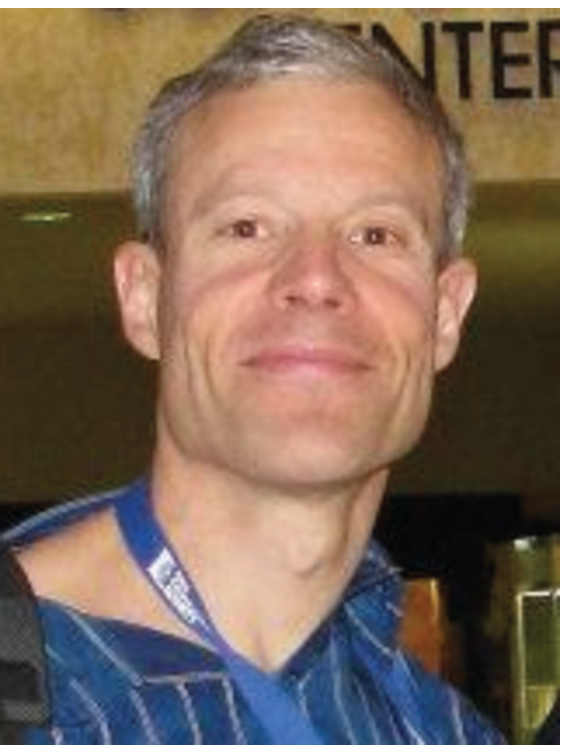}{Steven D. Blostein} received his B.S. degree in Electrical Engineering from Cornell University, Ithaca, NY and the M.S. and Ph.D. degrees in Electrical and 
Computer Engineering from the University of Illinois, Urbana-Champaign. He has been with the Department of Electrical and Computer 
Engineering Queen's University since 1988.  He was Multi-Rate Wireless Data Access Major Project leader in the Canadian Institute for Telecommunications Research, a consultant to industry and government in image compression, target tracking, radar imaging and wireless communications, and spent sabbatical leaves at Lockheed Martin Electronic Systems and at Communications Research Centre of Industry Canada. His interests lie in the application of signal processing to wireless communications systems, including synchronization, network MIMO and physical layer optimization for multimedia transmission.  He has been a member of the Samsung 4G Wireless Forum and an invited distinguished speaker.  He served as Department Head, IEEE Kingston Section Chair, Biennial Symposium on Communications Chair, and Editor for IEEE Transactions on Image Processing and IEEE Transactions on Wireless Communications.  He is a registered Professional Engineer in Ontario. 
\end{biography}

\epsfysize=3.2cm
\begin{biography}{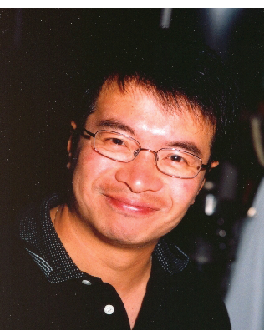}{Wai-Yip Chan}, also known as Geoffrey Chan, received his B.Eng. and M.Eng. degrees from Carleton University, Ottawa, and his Ph.D. degree from University of California, Santa Barbara, all in Electrical Engineering. He is currently with the Department of Electrical and Computer Engineering, Queen's University, Canada. He has held positions with the Communications Research Centre, Bell Northern Research (Nortel), McGill University, and Illinois Institute of Technology. His research interests are in speech processing and multimedia coding and communications. He is an associate editor of IEEE/ACM Transactions on Audio, Speech, and Language Processing, and of EURASIP Journal on Audio, Speech, and Music Processing. He has helped organize IEEE sponsored conferences on speech coding, image processing, and communications. He held a CAREER Award from the U.S. National Science Foundation.
\end{biography}

\end{document}